\documentclass[sigconf]{acmart}

\usepackage{amsfonts}       % blackboard math symbols
\usepackage{amsmath}
\usepackage{amssymb}
\usepackage{amsthm}
\usepackage{bm}
\usepackage{booktabs}       % professional-quality tables
\usepackage[T1]{fontenc}    % use 8-bit T1 fonts
\usepackage{graphicx}  %Required
\usepackage{hyperref}       % hyperlinks
\usepackage[utf8]{inputenc} % allow utf-8 input
\usepackage{microtype}      % microtypography
\usepackage{multirow}
\usepackage{nicefrac}       % compact symbols for 1/2, etc.
\usepackage{subcaption}% http://ctan.org/pkg/subcaption
\usepackage{url}            % simple URL typesetting\usepackage{}
\copyrightyear{2018}
\acmYear{2018}
\setcopyright{acmcopyright}
\acmConference[CIKM '18]{2018 ACM Conference on Information and Knowledge Management}{October 22--26, 2018}{Torino, Italy}
\acmBooktitle{2018 ACM Conference on Information and Knowledge Management (CIKM'18), October 22--26, 2018, Torino, Italy}
\acmPrice{15.00}
\acmDOI{10.1145/3269206.3271684} % fp0247
\acmISBN{978-1-4503-6014-2/18/10}
\begin{document}
\title{CoNet: Collaborative Cross Networks for Cross-Domain Recommendation}
%\title{Deep Transfer Learning for Recommender Systems}
%\title{Deep Transfer Learning for Cross-Domain Recommendation}
%\title{CoNet： Transfer Learning for Deep Recommender Systems}
\author{Guangneng Hu, Yu Zhang, Qiang Yang}
\affiliation{Department of Computer Science and Engineering, Hong Kong University of Science and Technology}
\email{njuhgn@gmail.com, yu.zhang.ust@gmail.com, qyang@cse.ust.hk}
%\email{username@institution.com}
% The default list of authors is too long for headers}
%\renewcommand{\shortauthors}{B. Trovato et al.}

\begin{abstract}
The cross-domain recommendation technique is an effective way of alleviating the data sparse issue in recommender systems by leveraging the knowledge from relevant domains. Transfer learning is a class of algorithms underlying these techniques. In this paper, we propose a novel transfer learning approach for cross-domain recommendation by using neural networks as the base model. In contrast to the matrix factorization based cross-domain techniques, our method is deep transfer learning, which can learn complex user-item interaction relationships. We assume that hidden layers in two base networks are connected by cross mappings, leading to the collaborative cross networks (CoNet). CoNet enables dual knowledge transfer across domains by introducing cross connections from one base network to another and vice versa. CoNet is achieved in multi-layer feedforward networks by adding dual connections and joint loss functions, which can be trained efficiently by back-propagation. The proposed model is thoroughly evaluated on two large real-world datasets. It outperforms baselines by relative improvements of 7.84\% in NDCG. We demonstrate the necessity of adaptively selecting representations to transfer. Our model can reduce tens of thousands training examples comparing with non-transfer methods and still has the competitive performance with them.
\end{abstract}

\keywords{Recommender Systems; Collaborative Filtering; Transfer Learning; Deep Learning}

\maketitle

\section{Introduction}

Collaborative filtering (CF) approaches, which model the preference of users on items based on their past interactions such as product ratings, are the corner stone for recommender systems. Matrix factorization (MF) is a class of CF methods which learn user latent factors and item latent factors by factorizing their interaction matrix~\cite{PMF,koren:2009}. Neural collaborative filtering is another class of CF methods which use neural networks to learn the complex user-item interaction function~\cite{dziugaite2015neural,cheng2016wide,he2017neural}. Neural networks have the ability to learn highly nonlinear function, which is suitable to learn the complex user-item interaction. Both traditional matrix factorization and neural collaborative filtering, however, suffer from the cold-start and data sparse issues.

One effective solution is to transfer the knowledge from relevant domains and the cross-domain recommendation techniques address such problems~\cite{berkovsky2007cross,li2009can,pan2011transfer,cantador2015cross}. In real life, a user typically participates several systems to acquire different information services. For example, a user installs applications in an app store and reads news from a website at the same time. It brings us an opportunity to improve the recommendation performance in the target service (or all services) by learning across domains. Following the above example, we can represent the app installation feedback using a binary matrix where the entries indicate whether a user has installed an app. Similarly, we use another binary matrix to indicate whether a user has read a news article. Typically these two matrices are highly sparse, and it is beneficial to learn them simultaneously. This idea is sharpened into the collective matrix factorization (CMF)~\cite{singh:2008} approach which jointly factorizes these two matrices by sharing the user latent factors. It combines CF on a target domain and another CF on an auxiliary domain, enabling knowledge transfer~\cite{pan:2008,zhang2017survey}. CMF, however, is a shallow model and has the difficulty in learning the complex user-item interaction function~\cite{dziugaite2015neural,he2017neural}. Moreover, its knowledge sharing is only limited in the lower level of user latent factors.

Motivated by benefitting from both knowledge transfer learning and learning interaction function, we propose a novel deep transfer learning approach for cross-domain recommendation using neural networks as the base model. Though neural CF approaches are proposed for single domain recommendation~\cite{he2017neural}, there are few related works to study knowledge transfer learning for cross-domain recommendation using neural networks. Instead, neural networks have been used as the base model in natural language processing~\cite{collobert2008unified,yang2017transfer} and computer vision~\cite{yosinski2014transferable,misra2016cross,doersch2017multi}. We explore how to use a neural network as the base model for each domain and enable the knowledge transfer on the entire network across domains. Then a few questions and challenges are raised: 1) What to transfer/share between these individual networks for each domain? 2) How to transfer/share during the learning of these individual networks for each domain? and 3) How is the performance compared with single domain neural learning and shallow cross-domain models?

This paper aims at proposing a novel deep transfer learning approach by answering these questions under cross-domain recommendation scenario. The usual transfer learning approach is to train a base network and then copy its first several layers to the corresponding first layers of a target network with fine-tuning or parameter frozen~\cite{yosinski2014transferable}. This way of transferring has possibly two weak points. Firstly, the shared-layer assumption is strong in practice as we find that it does not work well on real-world cross-domain datasets. Secondly, the knowledge transfer happens in one direction, i.e., only from source to target. Instead, we assume that hidden layers in two base networks are connected by dual mappings, which do not require them to be identical. We enable dual knowledge transfer across domains by introducing cross connections from one base network to another and vice versa, letting them benefit from each other. These ideas are sharpened into the proposed collaborative cross networks (CoNet). CoNet is achieved in simple multi-layer feedforward networks by using dual shortcut connections and joint loss functions, which can be trained efficiently by back-propagation.

The paper is organized as follows. We firstly introduce the preliminaries in Section~\ref{paper:preliminary}, including notations and the base network. In Section~\ref{paper:cross-stitich}, we then present an intuitive model to realize the cross-domain recommendation and point out several intrinsic weaknesses which limit its use. We propose a novel deep transfer learning approach for cross-domain recommendation, named collaborative cross networks (CoNet) in Section~\ref{paper:CoNet}. The core component is the cross connection units which enable knowledge transfer between source and target networks (Sec.~\ref{paper:cross-unit}). Its adaptive variant enforces the sparse structure which adaptively controls when to transfer (Sec.~\ref{paper:sparse-model}). In Section~\ref{paper:exp}, we experimentally show the benefits of both transfer learning and deep learning for improving the recommendation performance in terms of ranking metrics (Sec.~\ref{exp:comparison}). We show the necessity of adaptively selecting representations to transfer (Sec.~\ref{exp:sparse}). We reduce tens of thousands training examples without performance degradation by comparing with non-transfer models (Sec.~\ref{exp:transfer}), which can be used to save the cost/labor of labelling data. We review related works in Section~\ref{related-work} and conclude the paper in Section~\ref{paper:conclusion}. %following related works in Section~\ref{paper:relatedwork}.

\section{Preliminary}\label{paper:preliminary}

We first give the notations and describe the problem setting (Sec.~\ref{paper:notation}). We then review a multi-layer feedforward neural network as the base network for collaborative filtering (Sec.~\ref{paper:base}).

\subsection{Notation}\label{paper:notation}

We are given two domains, a source domain $\mathcal{S}$ (e.g., news recommendation) and a target domain $\mathcal{T}$ (e.g., app recommendation). As a running example, we let app recommendation be the target domain and news recommendation be the source domain. The set of users in both domains are shared, denoted by $\mathcal{U}$ (of size $m=|\mathcal{U}|$). Denote the set of items in $\mathcal{S}$ and $\mathcal{T}$ by $\mathcal{I}_S$ and $\mathcal{I}_T$ (of size $n_S=|\mathcal{I}_S|$ and $n_T=|\mathcal{I}_T|$), respectively. Each domain is a problem of collaborative filtering for implicit feedback~\cite{pan:2008,hu:2008}. For the target domain, let a binary matrix $\bm{R}_T \in \mathbb{R}^{m \times n_T}$ describe user-app installing interactions, where an entry $r_{ui} \in \{0,1\}$ is 1 (observed entries) if user $u$ has an interaction with app $i$ and 0 (unobserved) otherwise. Similarly, for the source domain, let another binary matrix $\bm{R}_S \in \mathbb{R}^{m \times n_S}$ describe user-news reading interactions, where the entry $r_{uj} \in \{0,1\}$ is 1 if user $u$ has an interaction with news $j$ and 0 otherwise. Usually the interaction matrix is very sparse since a user only consumed a very small subset of all items.

For the task of item recommendation, each user is only interested in identifying top-N items. The items are ranked by their predicted scores:
\begin{equation}
\hat r_{ui} = f(u,i | \Theta),
\end{equation}
where $f$ is the interaction function and $\Theta$ are model parameters. For matrix factorization (MF) techniques, the match function is the fixed dot product:
\begin{equation}\label{eq:pred-mf}
\hat r_{ui} = \bm{P}_u^T \bm{Q}_i,
\end{equation}
and parameters are latent vectors of users and items $\Theta = \{\bm{P}, \bm{Q}\}$ where $\bm{P} \in \mathbb{R}^{m \times d},\bm{Q} \in \mathbb{R}^{n \times d}$ and $d$ is the dimension. For neural CF approaches, neural networks are used to parameterize function $f$ and learn it from interactions:
\begin{equation}\label{eq:pred-ncf}
f(\bm{x}_{ui} | \bm{P}, \bm{Q}, \theta_f) = \phi_o( \phi_L(...(\phi_1(\bm{x}_{ui}))...)),
\end{equation}
where the input $\bm{x}_{ui} = [\bm{P}^T\bm{x}_u, \bm{Q}^T\bm{x}_i]$ is merged from projections of the user and the item, and the projections are based on their one-hot encodings $\bm{x}_u \in \{0,1\}^m,\bm{x}_i \in \{0,1\}^n$ and embedding matrices $\bm{P} \in \mathbb{R}^{m \times d},\bm{Q} \in \mathbb{R}^{n \times d}$. The output and hidden layers are computed by $\phi_o$ and $\{\phi_l\}$ in a multilayer feedforward neural network (FFNN), and the connection weight matrices and biases are denoted by $\theta_f$.

In our transfer/multitask learning approach for cross-domain recommendation, each domain is modelled by a neural network and these networks are jointly learned to improve the performance through mutual knowledge transfer. We review the base network in the following subsection before introducing the proposed model.

\subsection{Base Network}\label{paper:base}

We adopt an FFNN as the base network to parameterize the interaction function (see Eq.(\ref{eq:pred-ncf})). The base network is similar to the Deep model in~\cite{DeepYoutube,cheng2016wide} and the MLP model in~\cite{he2017neural}. The base network, as shown in Figure~\ref{fig:illustration} (the gray part or the blue part), consists of four modules with the information flow from the input $(u,i)$ to the output $\hat{r}_{ui}$ as follows.

{\bf Input }: $(u,i) \rightarrow \bm{x}_u,\bm{x}_i$. This module encodes user-item interaction indices. We adopt the one-hot encoding. It takes user $u$ and item $i$, and maps them into one-hot encodings $\bm{x}_u \in \{0,1\}^{m}$ and $\bm{x}_i \in \{0,1\}^{n}$ where only the element corresponding to that index is 1 and all others are 0.

{\bf Embedding }: $\bm{x}_u,\bm{x}_i \rightarrow \bm{x}_{ui}$. This module embeds one-hot encodings into continuous representations via two embedding matrices and then merges them as $\bm{x}_{ui} = [\bm{P}^T\bm{x}_u, \bm{Q}^T\bm{x}_i]$ to be the input of successive hidden layers.

{\bf Hidden layers}: $\bm{x}_{ui} \rightsquigarrow \bm{z}_{ui}$. This module takes the continuous representations from the embedding module and then transforms, through multi-hop say $L$, to a final latent representation $\bm{z}_{ui}= \phi_L(...(\phi_1(\bm{x}_{ui})...)$. This module consists of multiple hidden layers to learn nonlinear interaction between users and items.

{\bf Output }: $\bm{z}_{ui} \rightarrow \hat r_{ui}$. This module predicts the score $\hat r_{ui}$ for the given user-item pair based on the representation $\bm{z}_{ui}$ from the last layer of multi-hop module. Since we focus on one-class collaborative filtering, the output is the probability that the input pair is a positive interaction. This can be achieved by a softmax layer:
\begin{equation}\label{eq:output}
\hat r_{ui} = \phi_o(\bm{z}_{ui}) = \frac{1}{1 + \exp(-\bm{h}^T \bm{z}_{ui})},
\end{equation}
where $\bm{h}$ is the parameter.

\section{Cross-stitch Networks}\label{paper:cross-stitich}

We first introduce an intuitive model to realize cross-domain recommendation using neural networks, and point out several intrinsic strong assumptions limiting its use, which inspire the design of our model in the next section.

\begin{figure}
\begin{subfigure}{.25\textwidth}
  \centering
  \includegraphics[height=1.8in,width=4.4cm]{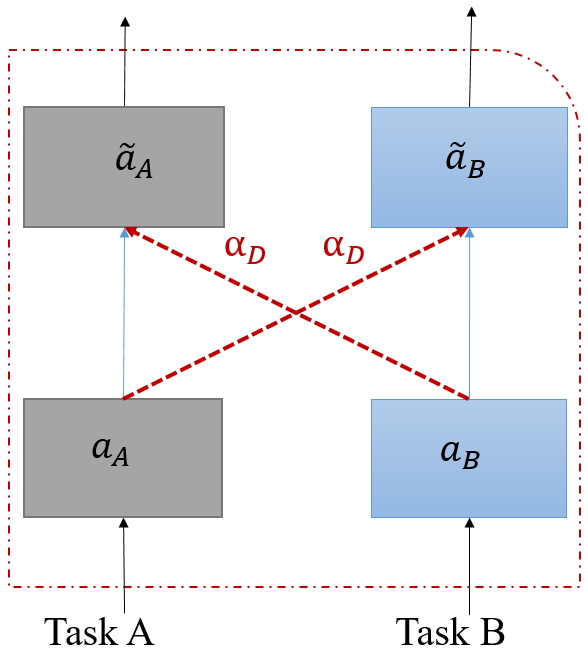} %{fig/cross-stitch-crop.pdf}
  \caption{Cross-stitch unit}
  \label{fig:cross-stitch}
\end{subfigure}%
\begin{subfigure}{.25\textwidth}
  \centering
  \includegraphics[height=1.8in,width=4.4cm]{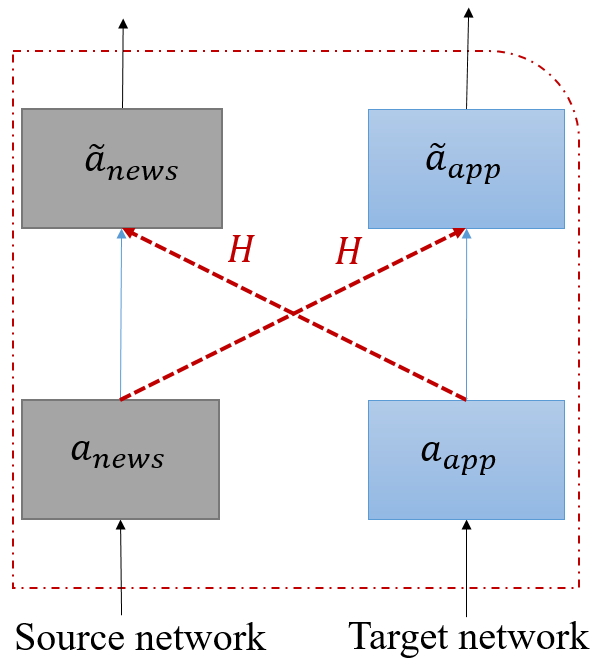} %{fig/cross-unit-crop.pdf}
  \caption{Cross connection unit}
  \label{fig:cross-unit}
\end{subfigure}
\caption{The cross-stitch unit~\cite{misra2016cross} (left) and the proposed cross connection unit (right). To enable knowledge transfer, the shared $\alpha_D$ in the cross-stitch network is a scalar while the shared $H$ in the proposed collaborative cross network is a matrix.}
\label{fig:cross-stitch-unit}
\end{figure}

Given two activation maps $a_A$ and $a_B$ from the $l$-th layer for two tasks $A$ and $B$, cross-stitch convolutional networks (CSN)~\cite{misra2016cross} learn linear combinations $\tilde{a}_A, \tilde{a}_B$ of both the input activations and feed these combinations as input to the successive layers' filters (see Fig.~\ref{fig:cross-stitch}):

\begin{equation}\label{eq:cross-stitch}
\tilde{a}_A^{ij}  = \alpha_S {a}_A^{ij} +  \alpha_D {a}_B^{ij}, \quad \tilde{a}_B^{ij}  = \alpha_S {a}_B^{ij} +  \alpha_D {a}_A^{ij},
\end{equation}
where the shared parameter $\alpha_D$ controls information shared/ transferred from the other network, $\alpha_S$ controls information from the task-specific network, and $(i,j)$ is the location in the activation map.

Although the cross-stitch unit indeed incorporates knowledge from the source domain (and target domain vice versa), there are several limitations of this simple stitch unit. Firstly, cross-stitch networks cannot process the case that the dimensions of contiguous layers are different. In other words, it assumes that the activations in successive layers are in the {\it same vector space}. This is not an issue in convolutional networks for computer vision since the activation maps of contiguous layers are in the same space~\cite{alexnet}. For collaborative filtering, however, it is not the case in typical multi-layer FFNNs where the architecture follows a tower pattern: the lower layers are wider and higher layers have smaller number of neurons~\cite{DeepYoutube,he2017neural}. Secondly, it assumes that the representations from other networks are {\it equally important} with weights being all the same scalar $\alpha_D$. Some features, however, are more useful and predictive and it should be learned attentively from data~\cite{visual-attention}. Thirdly, it assumes that the representations from other networks are {\it all useful} since it transfers activations from every location in a dense way. The sparse structure, however, plays a key role in general learning paradigm~\cite{doersch2017multi}. Instead, our model can be extended to learn the sparse structure on the task relationship matrices which are defined in Eq.~(\ref{eq:lasso}), with the help of the existing sparsity-induced regularization. As we will see in the experiments (see Table~\ref{tb:result} and Figure~\ref{fig:sparsity-mobile-amazon-80}), the sparse structure is necessary for generalization performance.

\section{Collaborative Cross Networks}\label{paper:CoNet}

To alleviate the limitations of the cross-stitch networks, we propose collaborative cross networks (CoNet) to transfer knowledge for the cross-domain recommendation. The core component is cross connection units (Sec.~\ref{paper:cross-unit}). Our cross unit generalizes the cross-stitch (Sec.~\ref{paper:basic-model}) and exploits the sparse structure (Sec.~\ref{paper:sparse-model}). We describe the model learning from implicit feedback datasets and the optimization process of the joint loss (Sec.~\ref{paper:learning}). A complexity analysis is also given (Sec.~\ref{paper:complexity}).

\subsection{Cross Connections Unit}\label{paper:cross-unit}

In this section, we present a novel soft-sharing approach for transferring knowledge for cross-domain recommendation. It relaxes the hard-sharing assumption~\cite{yosinski2014transferable} and is motivated by the cross-stitch networks~\cite{misra2016cross}.

We now introduce the cross connections unit to enable dual knowledge transfer as shown in Fig.~\ref{fig:cross-unit}. The central idea is simple, using a matrix rather than a scalar to transfer. Similarly to the cross-stitch network, the target network receives information from the source network and vice versa. In detail, let $\bm{a}_{app}$ be the representations of the $l$-th hidden layer and $\tilde{\bm{a}}_{app}$ be the input to the $l+1$-th in the app network, respectively. Similarly, they are $\bm{a}_{news}$ and $\tilde{\bm{a}}_{news}$ in the news network. The cross unit implements as follows:
\begin{subequations}\label{eq:cross-unit}
\begin{equation}
\tilde{\bm{a}}_{app} = \bm{W}_{app} \bm{a}_{app} + \bm{H} \bm{a}_{news},
\end{equation}
\begin{equation}
\tilde{\bm{a}}_{news} = \bm{W}_{news} \bm{a}_{news} + \bm{H} \bm{a}_{app},
\end{equation}
\end{subequations}
where $\bm{W}_{app}$ and $\bm{W}_{news}$ are weight matrices, and the matrix $\bm{H}$ controls the information from news network to app network and vice versa. The knowledge transferring happens in two directions, from source to target and from target to source. We enable dual knowledge transfer across domains and let them benefit from each other. When target domain data is sparse, the target network can still learn a good representation from that of the source network through the cross connection units. It only needs to learn ``residual'' target representations with the reference of source representations, making the target task learning easier and hence alleviating the data sparse issues. The role of matrix $\bm{H}$ is similar to the scalar $\alpha_D$ in the sense of enabling knowledge transfer between domains.

We give a closer look at the matrix $\bm{H}$ for it can alleviate all three issues faced by cross-stitch unit. Firstly, the successive layers can be in {\it different vector space} (spaces with different dimensions) since the matrix $\bm{H}$ can be used to match their dimension. For example, if the $l$-th layer ($\bm{a}_{app}$ and $\bm{a}_{news}$) has dimension 128, and the $l+1$-th layer ($\tilde{\bm{a}}_{app}$ and $\tilde{\bm{a}}_{news}$) has dimension 64, then the matrix $\bm{H} \in \mathbb{R}^{64 \times 128}$. Secondly, the entries of $\bm{H}$ are learned from data. They are likely not to be all the same, showing that the {\it importances} of transferred representations are different for each neuron/position. Thirdly, we can enforce some prior on the matrix $\bm{H}$ to exploit the structure of the neural architecture. The sparse structure can be enforced to adaptively {\it select useful representations} to transfer. Based on the cross connection units, we propose the CoNet models in the following sections, including a basic model (Sec.~\ref{paper:basic-model}) and an adaptive variant (Sec.~\ref{paper:sparse-model}).

\subsection{Basic Model}\label{paper:basic-model}

\begin{figure}
\centering
	\includegraphics[height=2.5in,width=7.6cm]{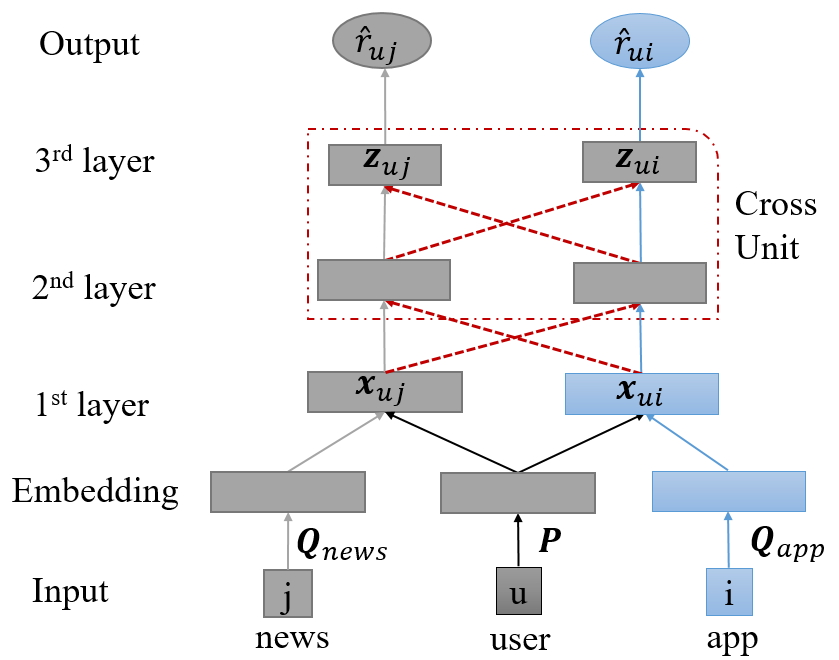} %{fig/illustration-crop.pdf}
	\caption{The proposed collaborative cross networks (CoNet) architecture (a version of three hidden layers and two cross units). We adopt a multilayer FFNN as the base network (grey or blue part, see Sec.~\ref{paper:base}). The red dotted lines indicate the cross connections which enable the dual knowledge transfer across domains. A cross unit is illustrated in the dotted rectangle box (see Fig.~\ref{fig:cross-unit}).}
	\label{fig:illustration}
\end{figure}

We propose the collaborative cross network (CoNet) model by adding cross connection units (see Sec.~\ref{paper:cross-unit}) and joint loss (see Eq.(\ref{eq:loss-conet})) to the entire FFNN, as shown in Figure~\ref{fig:illustration}. We firstly describe a basic model in this section and then present an adaptive variant in the next section.

We decompose the model parameters into two parts, task-shared and task-specific:
\begin{subequations}\label{eq:parameters}
\begin{equation}
\Theta_{app}  = \{\bm{P},(\bm{H}^l)_1^L\} \cup \{\bm{Q}_{app},{\theta_f}_{app}\}
\end{equation}
\begin{equation}
\Theta_{news} = \{\bm{P},(\bm{H}^l)_1^L\} \cup \{\bm{Q}_{news},{\theta_f}_{news}\},
\end{equation}
\end{subequations}
where $\bm{P}$ is the user embedding matrix and $\bm{Q}$ are the item embedding matrices with the subscript specifying the corresponding domain. The ${\theta_f} = \{(\bm{W}^l,b^l)_{l=1}^L, \bm{h}\}$ are the connection weight matrices and biases in the $L$-layer FFNN where $\bm{h}$ is the output weight as shown in Eq.(\ref{eq:output}). We stack the cross connections units on the top of the shared user embeddings, enabling deep knowledge transfer. Denote by $\bm{W}^l$ the weight matrix connecting from the $l$-th to the $l+1$-th layer (we ignore biases for simplicity), and by $\bm{H}^l$ the linear projection underlying the corresponding cross connections. Then two base networks are coupled by cross connections:
\begin{subequations}\label{eq:loss-cross}
\begin{equation}\label{eq:loss-cross-app}
 \bm{a}^{l+1}_{app} = \sigma(\bm{W}^l_{app} \bm{a}^l_{app} + \bm{H}^l \bm{a}_{news}^l),
\end{equation}
\begin{equation}\label{eq:loss-cross-news}
\bm{a}^{l+1}_{news} = \sigma(\bm{W}^l_{news} \bm{a}^l_{news} + \bm{H}^l \bm{a}_{app}^l),
\end{equation}
\end{subequations}
where the function $\sigma(\cdot)$ is the widely used rectified activation units (ReLU)~\cite{nair2010rectified}.
We can see that $\bm{a}^{l+1}_{app}$ receives two information flows: one is from the {\it transform} gate controlled by $\bm{W}^l_{app}$ and one is from the {\it transfer} gate controlled by $\bm{H}^l$ (similarly for the $\bm{a}^{l+1}_{news}$ in source network). We call $\bm{H}^l$ the relationship/transfer matrix since it learns to control how much sharing is needed. To reduce model parameters and make the model compact, we use the same linear transformation $\bm{H}^l$ for two directions, similar to the cross-stitch networks. Actually, using different matrices for two directions did not improve results on the evaluated datasets.

\subsection{Adaptive Model}\label{paper:sparse-model}

As we can see, the task relationship matrices $\{\bm{H}^l\}$ are crucial to the proposed CoNet model. We further enforce these matrices to have some structure. The assumption is that not all representations from another network are useful. We may expect that the representations coming from other domains are sparse and selective. This corresponds to enforcing a sparse prior on the structure and can be achieved by penalizing the task relationship matrix \{$\bm{H}^l$\} via some regularization. It may help the individual network to learn intrinsic representations for itself and other tasks. In other words, \{$\bm{H}^l$\} adaptively controls when to transfer.

We adopt the widely used sparsity-induced regularization---least absolute shrinkage and selection operator (lasso)~\cite{tibshirani1996regression}. In detail, denote by $r \times p$ the size of matrix $\bm{H}^l$ (usually $r = p/2$). That is, $\bm{H}^l$ linearly transforms representations $\bm{a}^l_{news} \in \mathbb{R}^p$ in the news network and the result is as part of the input to the next layer $\tilde{\bm{a}}^{l+1}_{app} \in \mathbb{R}^r$ in the app network (see Eq.(\ref{eq:loss-cross}) and Eq.(\ref{eq:cross-unit})). Denote by $h_{ij}$ the $(i,j)$ entry of $\bm{H}^l$. To induce overall sparsity, we impose the $\ell_1$-norm penalty on the entries $\{h_{ij}\}$ of $\bm{H}^l$:
\begin{equation}\label{eq:lasso}
\Omega(\bm{H}^l) = \lambda \sum\nolimits_{i=1}^r \sum\nolimits_{j=1}^p |h_{ij}|,
\end{equation}
where hyperparameter $\lambda$ controls the degree of sparsity. This corresponds to the lasso regularization. We call this sparse variant as the SCoNet model.

Other priors like low-rank ($\bm{H} \approx \bm{U}^T \bm{V}$) factorization are alternatives of sparse structure. And the lasso variants like group lasso and sparse group lasso are also possible. We adopt the general sparse prior and the widely used lasso regularization. The others are left for future work.

\subsection{Model Learning}\label{paper:learning}

Due to the nature of the implicit feedback and the task of item recommendation, the squared loss $(\hat r_{ui} - r_{ui})^2$ is not suitable since it is usually for rating regression/prediction. Instead, we adopt the cross-entropy loss:
\begin{equation}\label{eq:loss-base}
\mathcal{L}_0 =  - \sum\nolimits_{(u,i) \in \bm{R}^+ \cup \bm{R}^- } r_{ui} \log{\hat r_{ui}} + (1-r_{ui}) \log(1 - \hat r_{ui} ),
\end{equation}
where $\bm{R}^+$ and $\bm{R}^-$ are the observed interaction matrix and randomly sampled negative examples~\cite{pan:2008}, respectively. This objective function has probabilistic interpretation and is the negative logarithm likelihood of the following likelihood function:
\begin{equation}\label{eq:loss-likelihood}
    L(\Theta | \bm{R}^+ \cup \bm{R}^- ) = \prod\nolimits_{(u,i) \in \bm{R}^+} \hat r_{ui} \prod\nolimits_{(u,i) \in \bm{R}^-} (1 - \hat r_{ui}),
\end{equation}
where $\Theta$ are model parameters.

Now we define the joint loss function, leading to the proposed CoNet model which can be trained efficiently by back-propagation. Instantiating the {\it base loss} ($\mathcal{L}_0$) described
in Eq.~(\ref{eq:loss-base}) by the {\it loss of app} ($\mathcal{L}_{app}$) and {\it loss of news}
($\mathcal{L}_{news}$) recommendation, the objective function for the CoNet model is their joint losses:
\begin{equation}\label{eq:loss-conet}
%\mathcal{L}(\Theta) =  \alpha \mathcal{L}_{app} (\Theta_{app}) + (1-\alpha) \mathcal{L}_{news} (\Theta_{news}),
\mathcal{L}(\Theta) =  \mathcal{L}_{app} (\Theta_{app}) + \mathcal{L}_{news} (\Theta_{news}),
\end{equation}
where model parameters $\Theta =  \Theta_{app} \cup \Theta_{news}$. Note that $\Theta_{app}$ and $\Theta_{news}$ share user embeddings and transfer matrices $\{ \bm{P}, (\bm{H^l})_{l=1}^L \}$. For the CoNet-sparse model, the objective function is added by the term $\Omega(\bm{H}^l)$ in Eq.(\ref{eq:lasso}).

The objective function can be optimized by stochastic gradient descent (SGD) and its variants like adaptive moment method (Adam)~\cite{kingma2014adam}. The update equations are:
\begin{equation}
\Theta^{new} \leftarrow \Theta^{old} - \eta \frac{\partial L(\Theta)}{\partial \Theta},
\end{equation}
where $\eta$ is the learning rate. Typical deep learning library like TensorFlow\footnote{\url{https://www.tensorflow.org}} provides automatic differentiation and hence we omit the gradient equations $\frac{\partial L(\Theta)}{\partial \Theta}$ which can be computed by chain rule in back-propagation (BP).

\subsection{Complexity Analysis }\label{paper:complexity}

The model parameters $\Theta$ include $\{ \bm{P}, (\bm{H^l})_{l=1}^L \} \cup \{ \bm{Q}_{app},$ $ (\bm{W}^l_{app},b^l_{app})_{l=1}^L, \bm{h}_{app}\} \cup$  $\{ \bm{Q}_{news},(\bm{W}^l_{news},b^l_{news})_{l=1}^L$, \; $\bm{h}_{news}\}$, where the embedding matrices $\bm{P}$, $\bm{Q}_{app}$ and $\bm{Q}_{news}$ contain a large number of parameters since they depend on the input size of users and items. Typically, the number of neurons in a hidden layer is about one hundred. That is, the size of connection weight matrices and task relationship matrices is hundreds by hundreds. In total, the size of model parameters is linear with the input size and is close to the size of typical latent factors models~\cite{koren:2009} and neural CF approaches~\cite{he2017neural}.

During training, we update the target network using the target domain data and update the source network using the source domain data. The learning procedure is similar to the cross-stitch networks~\cite{misra2016cross}. And the cost of learning each base network is approximately equal to that of running a typical neural CF approach~\cite{he2017neural}. In total, the entire network can be efficiently trained by BP using mini-batch stochastic optimization.

\section{Experiment}\label{paper:exp}

We conduct thorough experiments to evaluate the proposed models. We show their superior performance over the state-of-the-art recommendation algorithms in a wide range of baselines (Sec.~\ref{exp:comparison}) and demonstrate the effectiveness of the sparse variant to select representations (Sec.~\ref{exp:sparse}). We quantify the benefit of knowledge transfer by reducing training examples  (Sec.~\ref{exp:transfer}). Furthermore, we conduct investigations on the sensitivity to hyperparameter (Sec.~\ref{exp:analysis}). We analyze the optimization efficiency (Sec.~\ref{exp:opt}) to help understand the proposed models.

\begin{table}[]					
\centering
\caption{Datasets and Statistics. }
\label{tb:data}			
\resizebox{0.49\textwidth}{!}{					
\begin{tabular}{c | c | ccc | ccc}		
\hline \hline											
{\multirow{2}{*}{Dataset}} & \multirow{2}{*}{\#Users} & \multicolumn{3}{c|}{Target Domain}  & \multicolumn{3}{c}{Source Domain}\\						
%\cline{3-9}						
\multicolumn{1}{c|}{}  &   &   \#Items &  \#Interactions  &  Density &   \#Items & \#Interactions & Density \\		
\hline \hline
Mobile &    23,111    & 14,348 & 1,164,394 & 0.351\%  & 29,921 & 617,146 & 0.089\% \\
\hline \hline
Amazon  &    80,763  & 93,799  & 1,323,101 & 0.017\% & 35,896 & 963,373 & 0.033\%  \\
\hline \hline
\end{tabular}	
}										
\end{table}

\subsection{Experimental Setup}

We begin the experiments by introducing the datasets, evaluation protocol, baselines, and implementation details.

\noindent
{\bf Dataset } We evaluate on two real-world cross-domain datasets. The first dataset, {\bf Mobile}\footnote{An anonymous version can be released later.}, is provided by a large internet company, i.e., Cheetah Mobile\footnote{\url{http://www.cmcm.com/en-us/}}. The information contains logs of user reading news, the history of app installation, and some metadata such as news publisher and user gender collected in one month in the US. The dataset we used contains 1,164,394 user-app installations and 617,146 user-news reading records. There are 23,111 shared users, 14,348 apps, and 29,921 news articles. We aim to improve the app recommendation by transferring knowledge from relevant news reading domain. The data sparsity is over 99.6\%.

The second dataset is a public {\bf Amazon} dataset\footnote{\url{http://snap.stanford.edu/data/web-Amazon.html}}, which has been widely used to evaluate the performance of collaborative filtering approaches~\cite{VBPR}. We use the two largest categories, Books and Movies \& TV, as the cross-domain. We convert the ratings of 4-5 as positive samples. The dataset we used contains 1,323,101 user-book ratings and 963,373 user-movie ratings. There are 80,763 shared users, 93,799 books, and 35,896 movies. We aim to improve the book recommendation by transferring knowledge from relevant movie watching domain. The data sparsity is over 99.9\%. The statistics are summarized in Table~\ref{tb:data}. As we can see, both datasets are very sparse and hence we hope improve performance by transferring knowledge from auxiliary domains.

\noindent
{\bf Evaluation Protocol } For item recommendation task, the leave-one-out (LOO) evaluation is widely used and we follow the protocol in~\cite{he2017neural}. That is, we reserve one interaction as the test item for each user. We determine hyper-parameters by randomly sampling another interaction per user as the validation/development set. We follow the common strategy which randomly samples 99 (negative) items that are not interacted by the user and then evaluate how well the recommender can rank the test item against these negative ones.

Since we aim at top-N item recommendation, the typical evaluation metrics are hit ratio (HR), normalized discounted cumulative gain (NDCG), and mean reciprocal rank (MRR), where the ranked list is cut off at $topN=10$. HR intuitively measures whether the reserved test item is present on the top-N list, defined as: $$HR = \frac{1}{|\mathcal{U}|} \sum\nolimits_{u \in \mathcal{U}} \delta(p_u \leq topN),$$
where $p_u$ is the hit position for the test item of user $u$, and $\delta(\cdot)$ is the indicator function. NDCG and MRR also account for the rank of the hit position, respectively defined as:
$$NDCG = \frac{1}{|\mathcal{U}|} \sum\nolimits_{u \in \mathcal{U}} \frac{\log2} {\log(p_u + 1)}, \; MRR = \frac{1}{|\mathcal{U}|} \sum\nolimits_{u \in \mathcal{U}} \frac{1}{p_u}.$$ A higher value indicates better performance.

\noindent
{\bf Baseline} We compare with various baselines:

\resizebox{0.45\textwidth}{!}{
\begin{tabular}{|c|c|c|}
\hline
 Baselines    & Shallow  method                            & Deep method      \\
\hline
Single-domain & BPRMF~\cite{rendle-etal:2009}              & MLP~\cite{he2017neural}  \\
\hline
Cross-domain  & CDCF~\cite{CDCF-FM}, CMF~\cite{singh:2008} & MLP++, CSN~\cite{misra2016cross} \\
\hline
\end{tabular}
}

{\it BPRMF}: Bayesian personalized ranking~\cite{rendle-etal:2009} is a typical latent factors CF approach which learns the user and item factors via matrix factorization and pairwise rank loss. It computes the prediction score by $\hat r_{ui} = \bm{P}_u^T \bm{Q}_i$ (see Eq.(\ref{eq:pred-mf})). It is a shallow model and learns on the target domain only.

{\it MLP}: Multilayer perceptron~\cite{he2017neural} is a typical neural CF approach which learns user-item interaction function using neural networks. MLP corresponds to the base network as described in Section~\ref{paper:base}. It is a deep model and learns on the target domain only.

{\it MLP++}: We combine two MLPs by sharing the user embedding matrix only. This is a degenerated CoNet which has no cross connection units. It is a simple/shallow knowledge transfer approach applied to two domains.

{\it CDCF}: Cross-Domain CF with factorization machines (FM)~\cite{CDCF-FM} is a state-of-the-art cross-domain recommendation which extends FM~\cite{rendle:2012}. It is a context-aware approach which applies factorization on the merged domains (aligned by the shared users). That is, the auxiliary domain is used as context. On the Mobile dataset, the context for a user in the target app domain is her history of reading news in the source news domain. Similarly, the context for a user in the target book domain is her history of watching movies in the source movie domain on the Amazon dataset. The feature vector for the input is a sparse vector $\bm{x} \in \mathbb{R}^{m+n_T+n_S}$ where the non-zero entries are as follows: 1) the index for user id, 2) the index for item id (target domain), and all indices for her reading articles/watching movies (source domain). It showed better performance than other cross-domain methods like triadic (tensor) factorization~\cite{triadic}. It is a shallow cross-domain model.

{\it CMF}: Collective matrix factorization~\cite{singh:2008} is a multi-relation learning approach which jointly factorizes matrices of individual domains. Here, the relation is user-item interaction. On Mobile, the two matrices are $\bm{A}=$ ``user by app'' and $\bm{B}=$ ``user by news'' respectively. Similarly, they are $\bm{A}=$ ``user by book'' and $\bm{B}=$ ``user by movie'' on Amazon. The shared user factors $\bm{P}$ enable knowledge transfer between two domains. Then CMF factorizes matrices $\bm{A}$ and $\bm{B}$ simultaneously by sharing the user latent factors: $\bm{A} \approx \bm{P}^T \bm{Q}_A$ and $\bm{B} \approx \bm{P}^T \bm{Q}_B$. It is a shallow model and jointly learns on two domains. This can be thought of a non-deep transfer/multitask learning approach for cross-domain recommendation.

{\it CSN}: The cross-stitch network method~\cite{misra2016cross}, described in Section~\ref{paper:cross-stitich}, is a good competitor. It is a deep multitask learning model which jointly learns two base networks. It enables knowledge transfer via a linear combination of activation maps from two networks via a shared coefficient, i.e., $\alpha_D$ in Eq.(\ref{eq:cross-stitch}). This is a deep transfer/multitask learning approach for cross-domain recommendation.

\noindent
{\bf Implementation } For BPRMF, we use LightFM's implementation\footnote{\url{https://github.com/lyst/lightfm}} which is a popular CF library. For CDCF, we adapt the official libFM implementation\footnote{\url{http://www.libfm.org}}. For MLP, we use the code released by its authors\footnote{\url{https://github.com/hexiangnan/neural_collaborative_filtering}}. For CMF, we use a Python version reference to the original Matlab code\footnote{\url{http://www.cs.cmu.edu/~ajit/cmf/}}. Our methods are implemented using TensorFlow. Parameters are randomly initialized from Gaussian $\mathcal{N}(0, 0.01^2)$. The optimizer is Adam with initial learning rate 0.001. The size of mini batch is 128. The ratio of negative sampling is 1. As for the design of network structure, we adopt a tower pattern, halving the layer size for each successive higher layer. Specifically, the configuration of hidden layers in the base network is $[64 \rightarrow 32 \rightarrow 16 \rightarrow 8]$. This is also the network configuration of the MLP model. For CSN, it requires that the number of neurons in each hidden layer is the same. The configuration notation $[64] * 4$ equals $[64 \rightarrow 64 \rightarrow 64 \rightarrow 64]$. We investigate several typical configurations.

\subsection{Comparing Different Approaches}\label{exp:comparison}

\begin{table*}[]
\centering
\caption{Comparison results of different methods on two datasets. The best results are boldfaced and the best baselines are marked with stars.}
\label{tb:result}
\resizebox{0.9\textwidth}{!}{
\begin{tabular}{ cc | ccc ccc cc | c c}
\hline \hline
Dataset & Metric & BPRMF &  CMF   & CDCF  & MLP   & MLP++ & CSN   & CoNet & SCoNet & improve& paired $t$-test \\
\hline \hline
        & HR     & .6175 & .7879  & .7812 & .8405 & .8445 &.8458* & .8480 & {\bf .8583} & 1.47\% & $p=0.20$\\
\cline{2-12}
Mobile  & NDCG   & .4891 & .5740  & .5875 & .6615 & .6683 &.6733* & .6754 & {\bf .6887} & 2.29\% & $p=0.25$\\	
\cline{2-12}
        & MRR    & .4489 & .5067  & .5265 & .6210 & .6268 &.6366* & .6373 & {\bf .6475} & 1.71\% & $p=0.34$\\
\hline \hline
        & HR     & .4723 & .3712  & .3685 & .5014 & .5050*& .4962 & .5167 & {\bf .5338} & 5.70\% & $p=0.02$\\
\cline{2-12}
Amazon  & NDCG   & .3016 & .2378  & .2307 & .3143 & .3175*& .3068 & .3261 & {\bf .3424} & 7.84\% & $p=0.03$\\	
\cline{2-12}
        & MRR    & .2971 & .1966  & .1884 & .3113*& .3053 & .2964 & .3163 & {\bf .3351} & 7.65\% & $p=0.05$\\
\hline \hline
\end{tabular}
}
\end{table*}

In this section, we report the recommendation performance of different methods and discuss the findings. Table~\ref{tb:result} shows the results of different models on the two datasets under three ranking metrics. The last two columns are the relative improvement and its paired $t$-test of our model vs. the best baselines.
%http://www.socscistatistics.com/tests/studentttest/Default2.aspx
We can see that our proposed neural models are better than the base network (MLP), the shallow cross-domain models (CMF and CDCF) learned using two domains information, and the deep cross-domain model (MLP++ and CSN) on both datasets.

On Mobile, our model achieves 4.28\% improvements in terms of MRR comparing with the non-transfer MLP, showing the benefits of knowledge transfer. Note that, the way of pre-training an MLP on source domain and then transferring user embeddings to target domain as warm-up did not achieve much improvement. In fact, the improvement is so small that it can be ignored. It shows the necessity of dual knowledge transfer in a deep way. Our model improves more than 20\% in terms of MRR comparing with CDCF and CMF, showing the effectiveness of deep neural approaches. Together, our neural models consistently give better performance than other existing methods. Within our models (SCoNet vs CoNet), enforcing sparse structure on the task relationship matrices are useful. Note that, the dropout technique and $\ell_2$ norm penalty did not achieve these improvements. They may harm the performance in some cases. It shows the necessity of selecting representations.

On Amazon, our model achieves 7.84\% improvements in terms of NDCG comparing with the best baselines (MLP++), showing the benefits of knowledge transfer. Compared to the BPRMF, the inferior performance of CMF and CDCF shows the difficulty in transferring knowledge between Amazon Books and Movies, but our models also achieve good results. Comparing MLP++ and MLP, sharing user embedding is sightly better than the base network due to shallow knowledge transfer. Within our models, enforcing sparse structure on the task relationship matrices are also useful.

CSN is inferior to the proposed CoNet models on both datasets. Moreover, it is surprising that the CSN has some difficulty in benefitting from knowledge transfer on the Amazon dataset since it is inferior to the non-transfer base network MLP. The reason is possibly that the assumptions of CSN are not appropriate: all representations from the auxiliary domain are {\it equally} important and are {\it all} useful. By using a matrix $\bm{H}$ rather than a scalar $\alpha_D$, we can relax the first assumption. And by enforcing a sparse structure on the matrix, we also relax the second assumption.

Note that the relative improvement of the proposed model vs. the best baseline is more significant on the Amazon dataset than that on the Mobile dataset, though the Amazon is much sparser than the Mobile (see Table~\ref{tb:data}). One explanation is that the relatedness the book and movie domains is much larger than that between the app and news domains. This will benefit all cross-domain methods including CMF, CDCF, and CSN, since they exploit information from both two domains. Another possibility is that the noise from auxiliary domain proposes a challenge for transferring knowledge. This shows that the proposed model is more effective since it can select useful representations from the source network and ignore the noisy ones. In the next section, we give a closer look at the impact of the sparse structure.

\subsection{Impact of Sparsity: Selecting Representations to Transfer}\label{exp:sparse}

On two real-world datasets, it both shows the usefulness of enforcing sparse structure on the task relationship matrices $\bm{H}$. We now quantify the contributions of the sparsity to CoNet. We investigate the impact of the sparsity by controlling the difference of architectures between CSN and CoNet. That is, we let them have the same architecture configuration. As a consequence, the performance of ablation comes from different means of knowledge transfer: scalar $\alpha_D$ used in CSN and sparse matrix $\bm{H}$ used in SCoNet.

\begin{figure*}
\centering
\begin{subfigure}{.248\textwidth}
  \centering
  \includegraphics[height=1.3in,width=4.6cm]{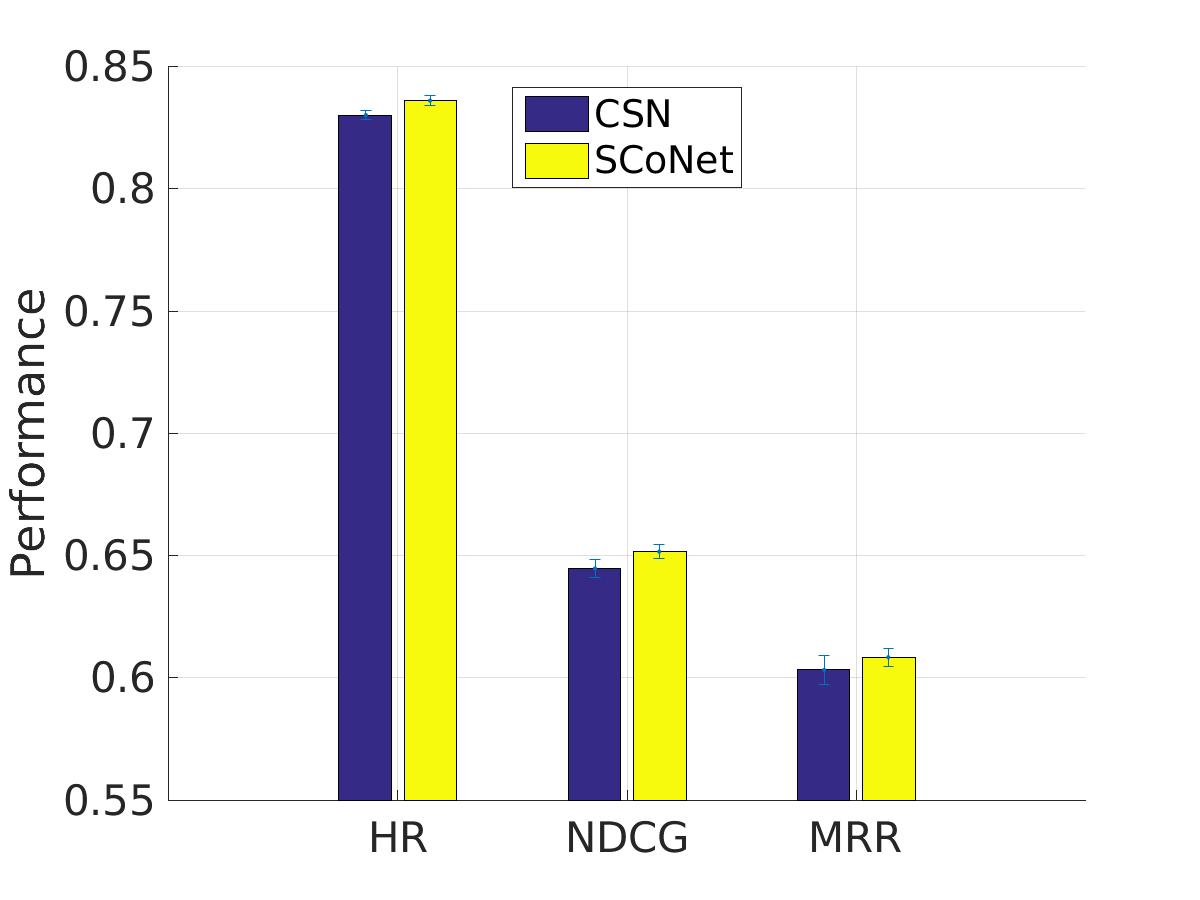} %{fig/sparsity-mobile-his-16-crop.pdf}
  %\caption{$[16] * 4$}
  \label{fig:sfig1}
\end{subfigure}%
\begin{subfigure}{.248\textwidth}
  \centering
  \includegraphics[height=1.3in,width=4.6cm]{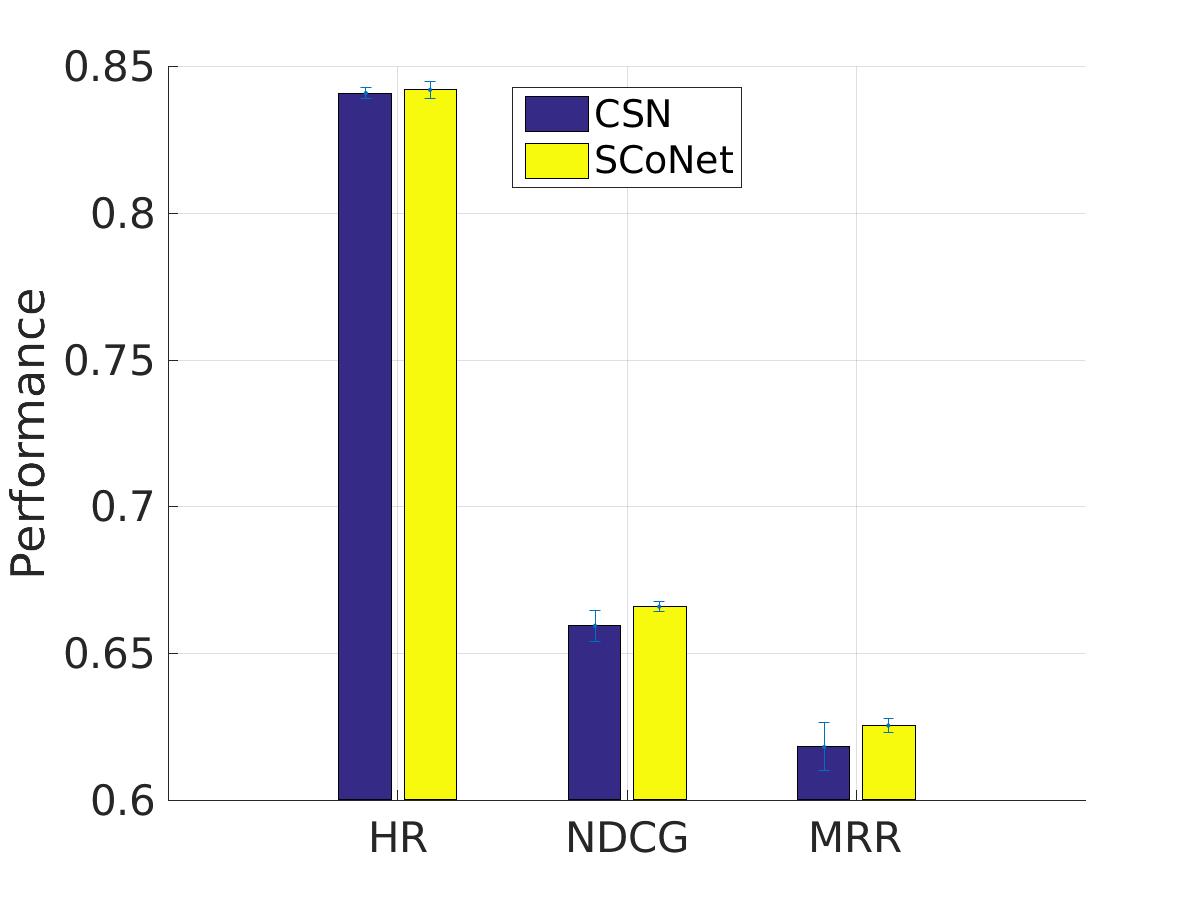}
  %\caption{$[32] * 4$}
  \label{fig:sfig2}
\end{subfigure}
\begin{subfigure}{.248\textwidth}
  \centering
  \includegraphics[height=1.3in,width=4.6cm]{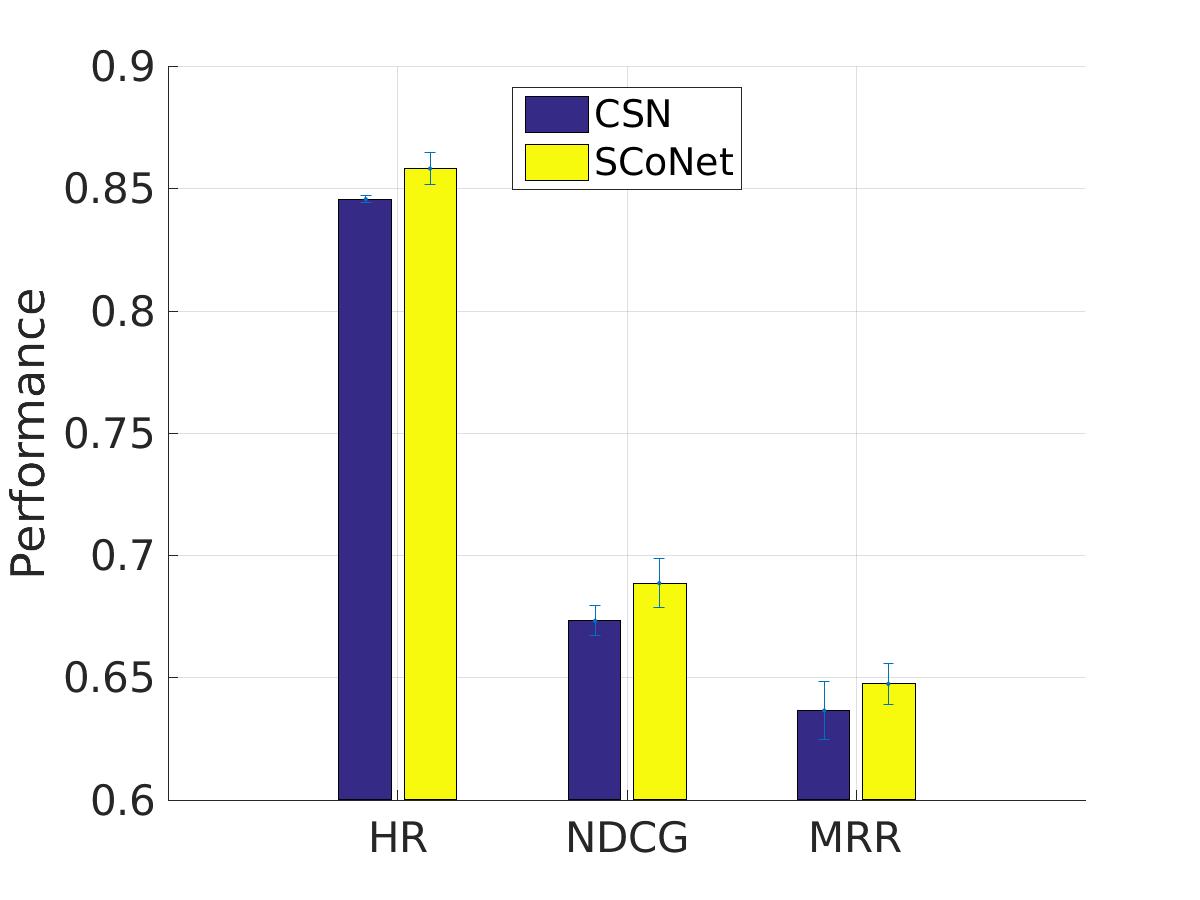}
  %\caption{$[64] * 4$}
  \label{fig:sfig1}
\end{subfigure}%
\begin{subfigure}{.248\textwidth}
  \centering
  \includegraphics[height=1.3in,width=4.6cm]{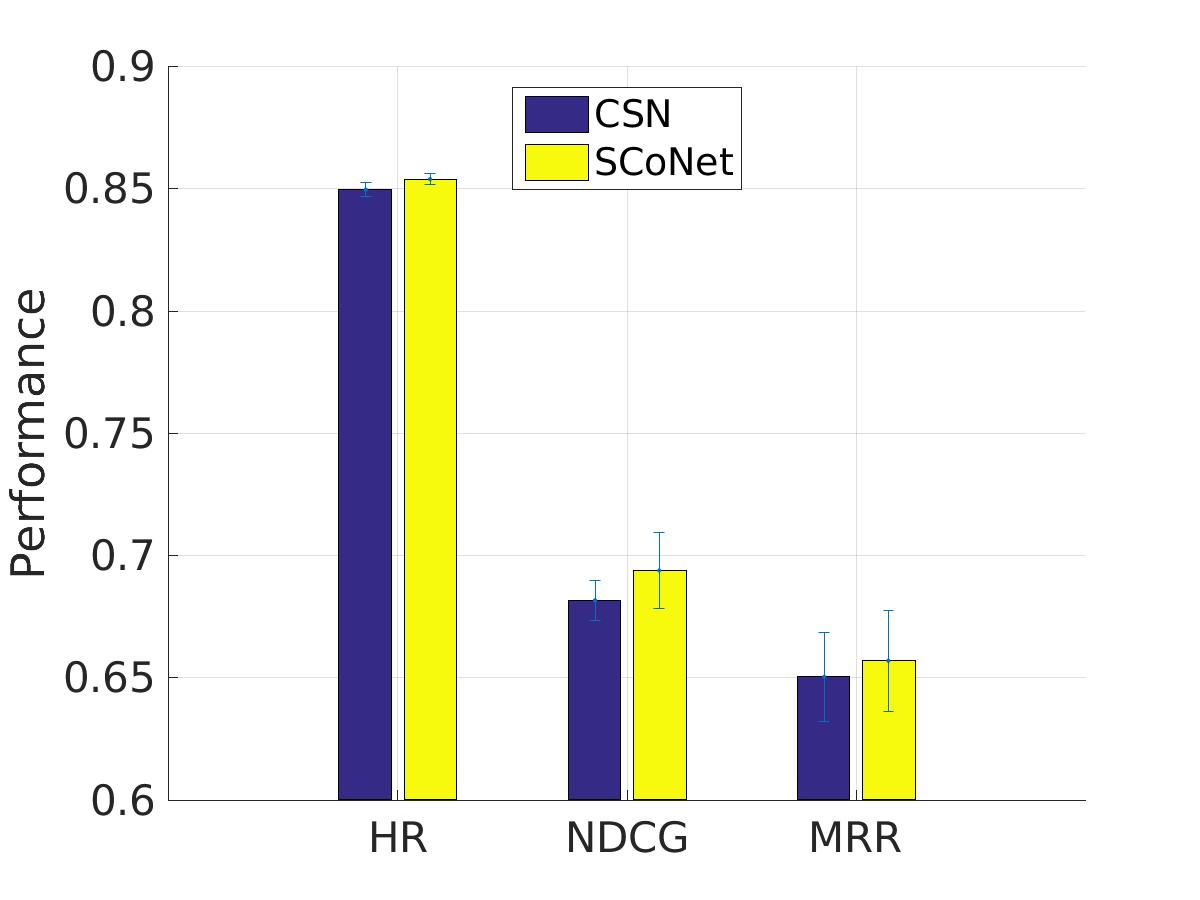}
  %\caption{$[80] * 4$}
  \label{fig:sfig2}
\end{subfigure}

\begin{subfigure}{.248\textwidth}
  \centering
  \includegraphics[height=1.3in,width=4.6cm]{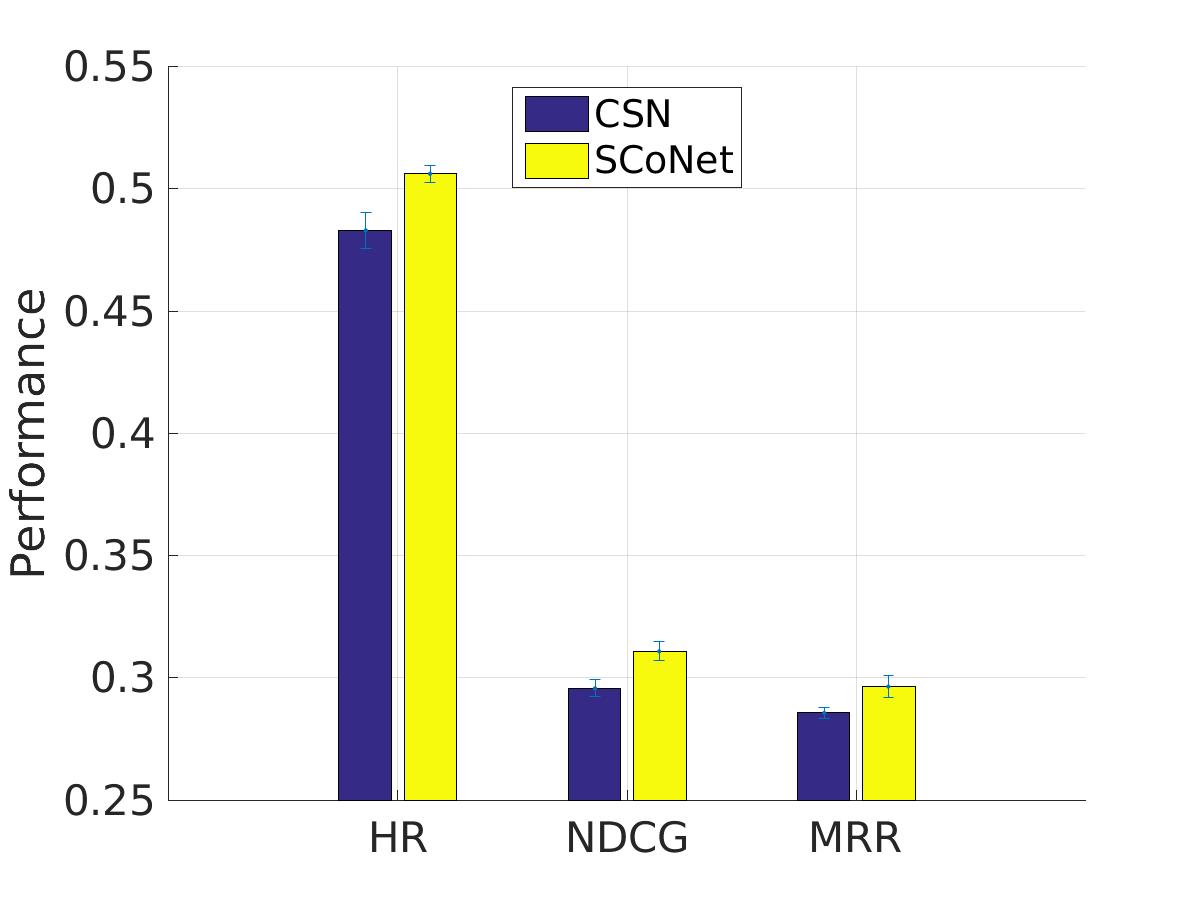}
  %\caption{$[16] * 4$}
  \label{fig:sfig1}
\end{subfigure}%
\begin{subfigure}{.248\textwidth}
  \centering
  \includegraphics[height=1.3in,width=4.6cm]{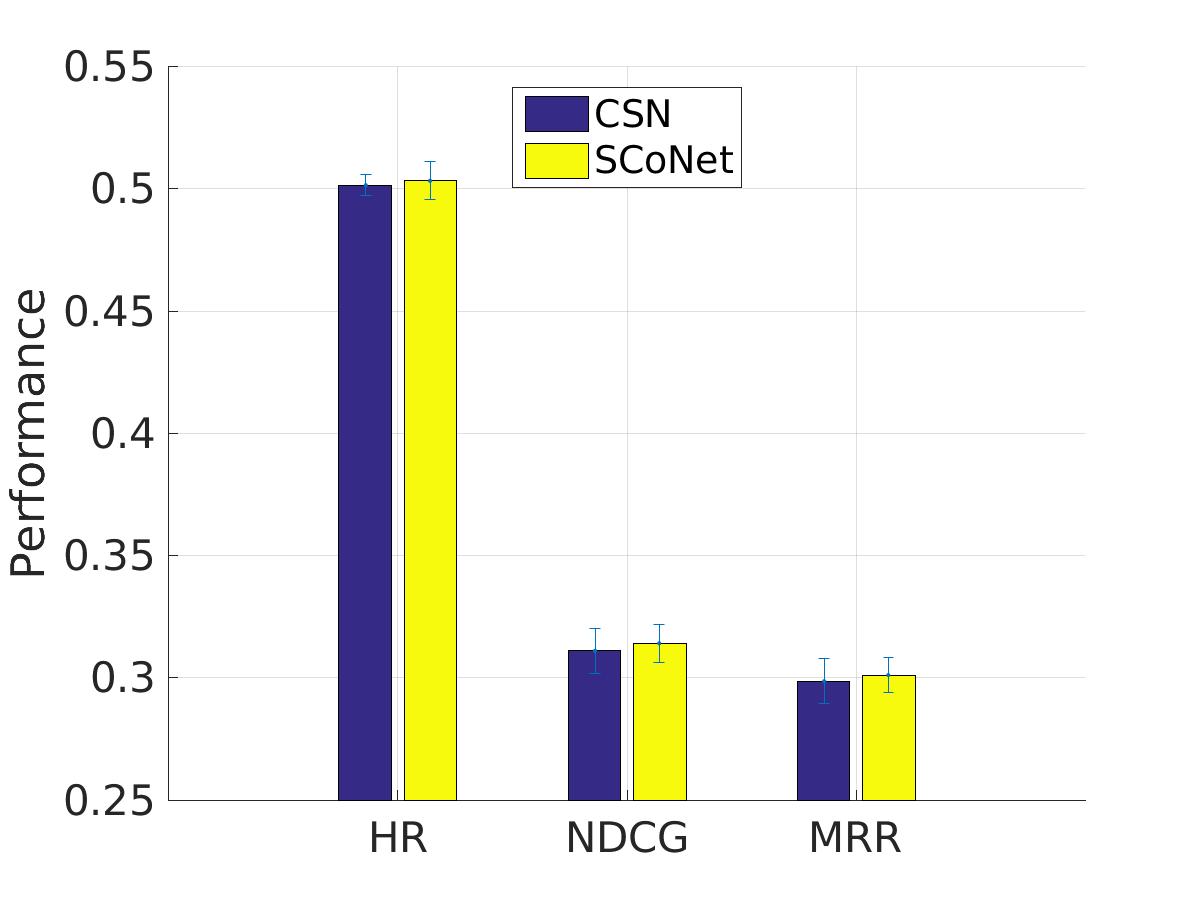}
  %\caption{$[32] * 4$}
  \label{fig:sfig2}
\end{subfigure}
\begin{subfigure}{.248\textwidth}
  \centering
  \includegraphics[height=1.3in,width=4.6cm]{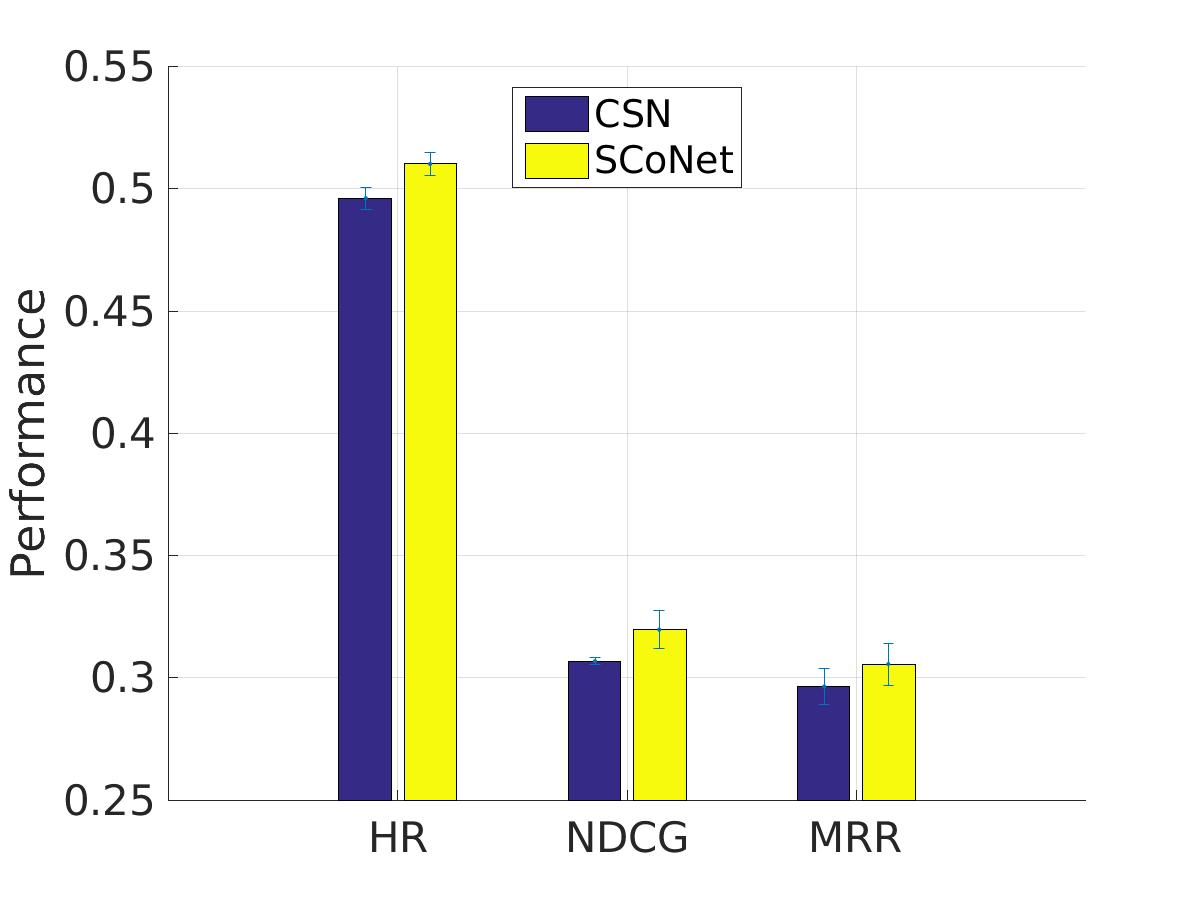}
  %\caption{$[64] * 4$}
  \label{fig:sfig1}
\end{subfigure}%
\begin{subfigure}{.248\textwidth}
  \centering
  \includegraphics[height=1.3in,width=4.6cm]{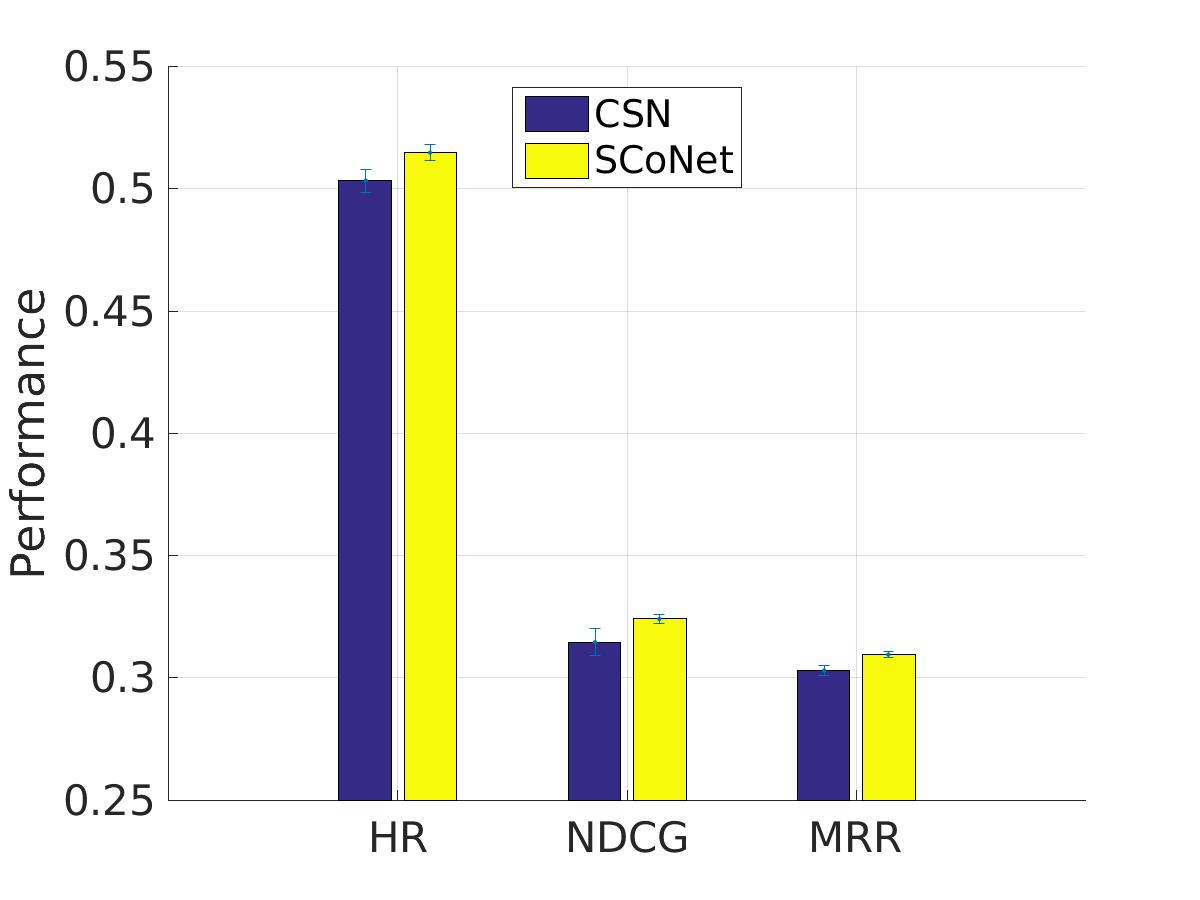}
  %\caption{$[80] * 4$}
  \label{fig:sfig2}
\end{subfigure}
\caption{Impact of the sparsity on the Mobile (top) and Amazon (bottom) datasets. From left to right, configurations are $\{16,32,64,80\} * 4$.}
\label{fig:sparsity-mobile-amazon-80}
\end{figure*}

Figure~\ref{fig:sparsity-mobile-amazon-80} shows the results on the Mobile and Amazon datasets under several typical architectures. We can see that the sparsity contributes to performance improvements and it is necessary to introduce the sparsity in general settings. On the Mobile data, introducing the sparsity improves the NDCG by relatively 2.29\%. On the Amazon data, introducing the sparsity improves the NDCG by relatively 4.21\%. These results show that it is beneficial to introduce the sparsity and to select representations to transfer on both datasets.

\subsection{Benefit of Transferring: Reducing Labelled Data}\label{exp:transfer}

Transfer learning can reduce the labor and cost of labelling data instances. In this section, we quantify the benefit of knowledge transfer by comparing with non-transfer methods. That is, we gradually reduce the number of training examples in the target domain until the performance of the proposed model is inferior to the non-transfer MLP model. The more training examples we can reduce, the more benefit we can get from transferring knowledge.

Referring to Table~\ref{tb:data}, there are about 50 examples per user on the Mobile dataset. We gradually reduce one and two training examples per user, respectively, to investigate the benefit of knowledge transfer. The results are shown in Table~\ref{tb:reduction-mobile-amazon} where the rows corresponding to reduction percentage 0\% are copied from Table~\ref{tb:result} for clarity. The number 2.05\% is approximately corresponding to reducing one training example per user. The results show that we can save the cost of labelling about $30,000$ training examples by transferring knowledge from the news domain but still have comparable performance with the MLP model, a non-transfer baseline.

According to Table~\ref{tb:data}, there are about 16 examples per user on the Amazon dataset. With a similar setting to the Mobile dataset, the results shown in Table~\ref{tb:reduction-mobile-amazon} indicates that we can save the cost of labelling about $20,000$ training examples by transferring knowledge from movie domain. Note that the Amazon dataset is extremely sparse (the density is only 0.017\%), implying that there is difficulty in acquiring many training examples. Under this scenario, our transfer models are an effective way of alleviating the issue of data sparsity and the cost of collecting data.

\begin{table}[]
\centering
\caption{The performance when reducing training examples. Results with stars are inferior to MLP.}
\label{tb:reduction-mobile-amazon}
\resizebox{0.45\textwidth}{!}{
\begin{tabular}{c| c| cc| ccc }
\hline \hline
\multirow{2}{*}{Dataset}  & \multirow{2}{*}{Method}      & \multicolumn{2}{c|}{Reduction} & \multirow{2}{*}{HR}  & \multirow{2}{*}{NDCG}  & \multirow{2}{*}{MRR}  \\
\cline{3-4}
  &                          & percent & amount &       &       &   \\
\hline \hline
\multirow{4}{*}{Mobile}& MLP & 0\%     & 0      & .8405 & .6615 & .6210    \\
\cline{2-7} \cline{2-7}
&\multirow{3}{*}{SCoNet}& 0\%     & 0      & .8547 & .6802 & .6431 \\
\cline{3-7}
&                            & 2.05\%  & 23,031 & .8439 & .6640 & .6238 \\
\cline{3-7}
&                            & 4.06\%  & 45,468 & .8347*& .6515*& .6115*\\
\hline \hline
\multirow{4}{*}{Amazon}& MLP & 0\%     & 0      & .5014 & .3143 & .3113    \\
\cline{2-7} \cline{2-7}
&\multirow{3}{*}{SCoNet}& 0\%     & 0      & .5338 & .3424 & .3351     \\
\cline{3-7}
&                            & 1.11\%  & 12,850 & .5110 & .3209 & .3080* \\
\cline{3-7}
&                            & 2.18\%  & 25,318 & .4946*& .3082*& .2968* \\
\hline \hline
\end{tabular}
}
\end{table}

\subsection{Sensitivity Analysis}\label{exp:analysis}

We analyze the sensitivity to the sparse penalty which controls the sparsity ($\lambda$ in Eq.(\ref{eq:lasso})). Results are shown on the Mobile data only due to space limit and we give the corresponding conclusions on the Amazon data. Figure~\ref{fig:lassopenalty} shows the performance varying with the penalty of sparsity enforcing on the task relationship matrices $\bm{H}$. On the Mobile data, the performance achieves good results at 0.1 (default) and 5.0, and it is 0.1 (default) and 1.0 on the Amazon data (not shown).

\subsection{Optimization Performance}\label{exp:opt}

\begin{figure*}
\centering
\begin{subfigure}{.33\textwidth}
  \centering
  \includegraphics[height=1.8in,width=5.6cm]{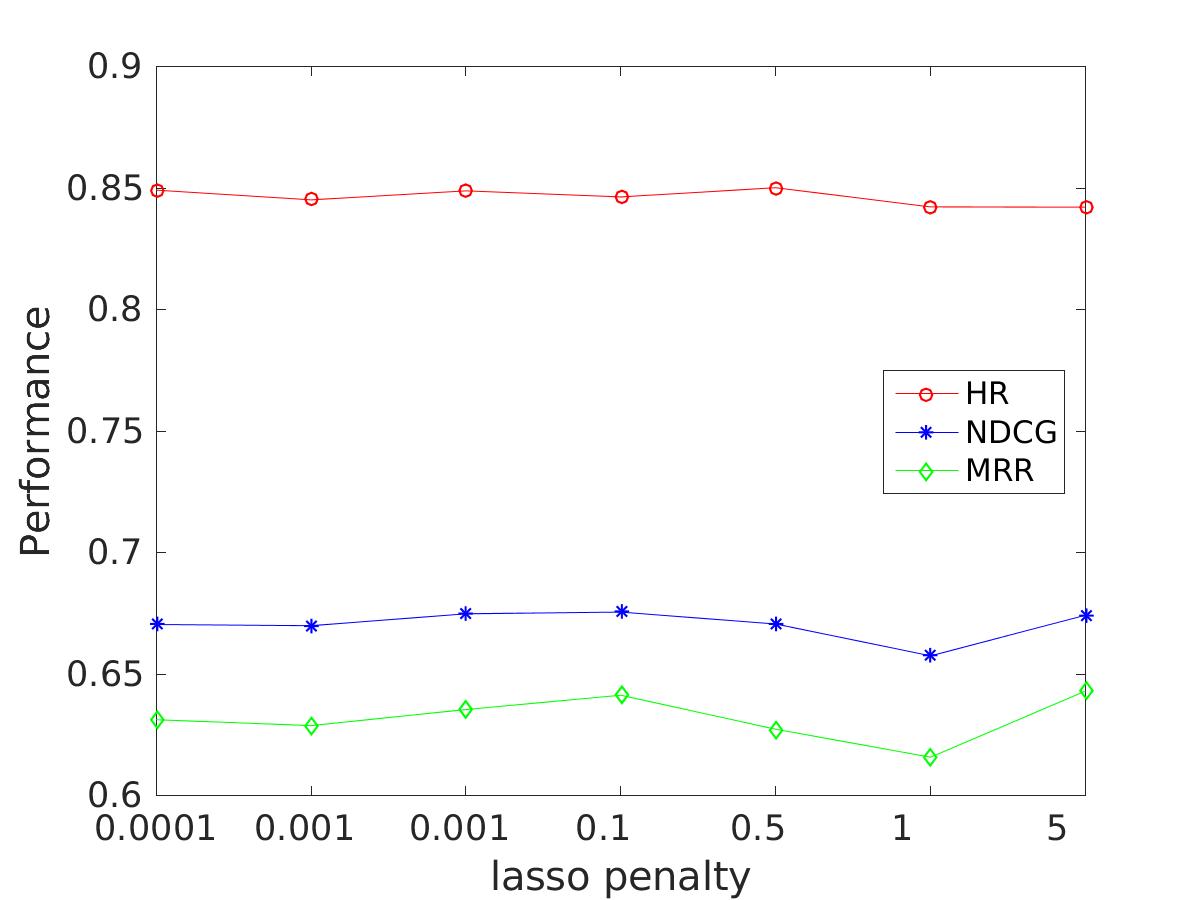} %{fig/lassopenalty-crop.pdf}
  \caption{Sparse penalty}
\label{fig:lassopenalty}
\end{subfigure}%
\begin{subfigure}{.33\textwidth}
  \centering
  \includegraphics[height=1.8in,width=5.7cm]{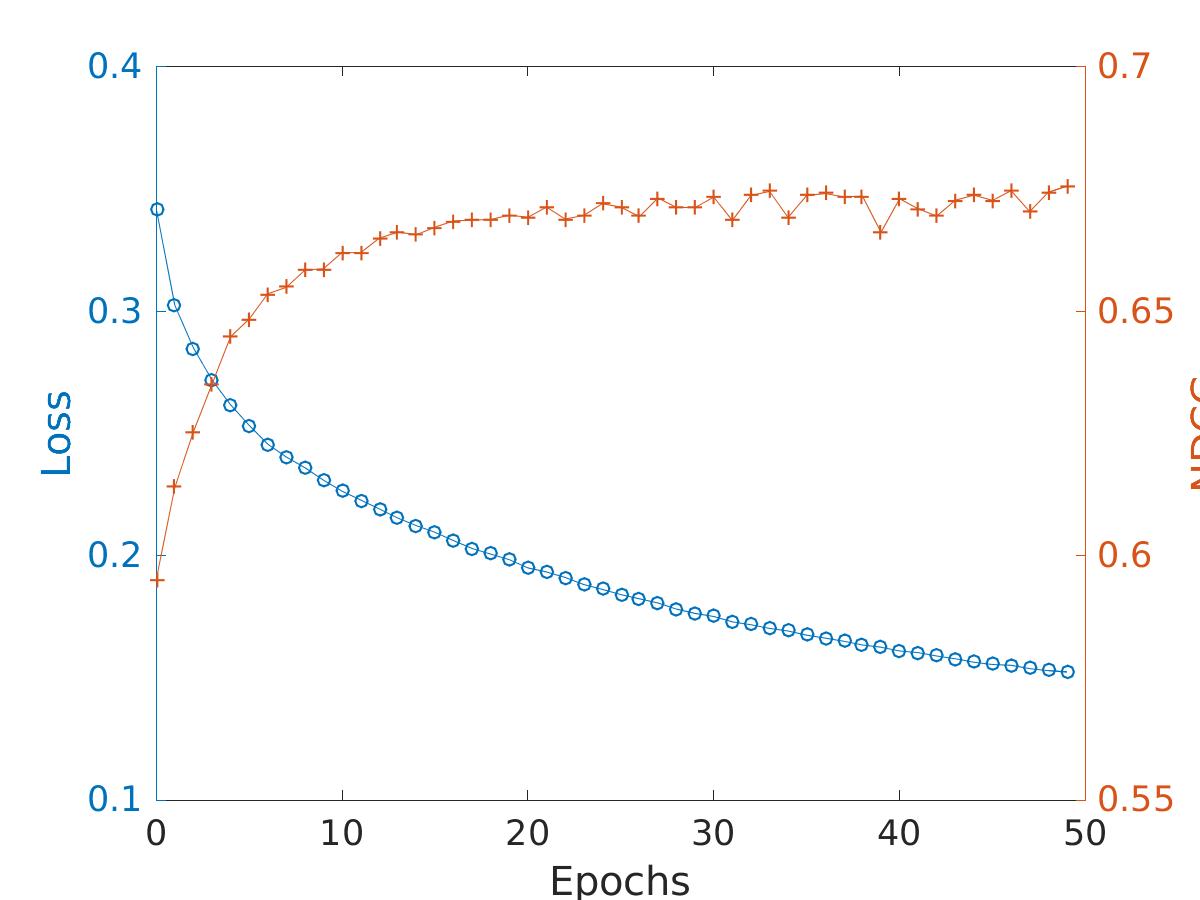} %{fig/loss-ndcg-crop.pdf}
  \caption{Loss and performance}
\label{fig:epoch-loss}
\end{subfigure}
\begin{subfigure}{.33\textwidth}
  \centering
  \includegraphics[height=1.8in,width=5.6cm]{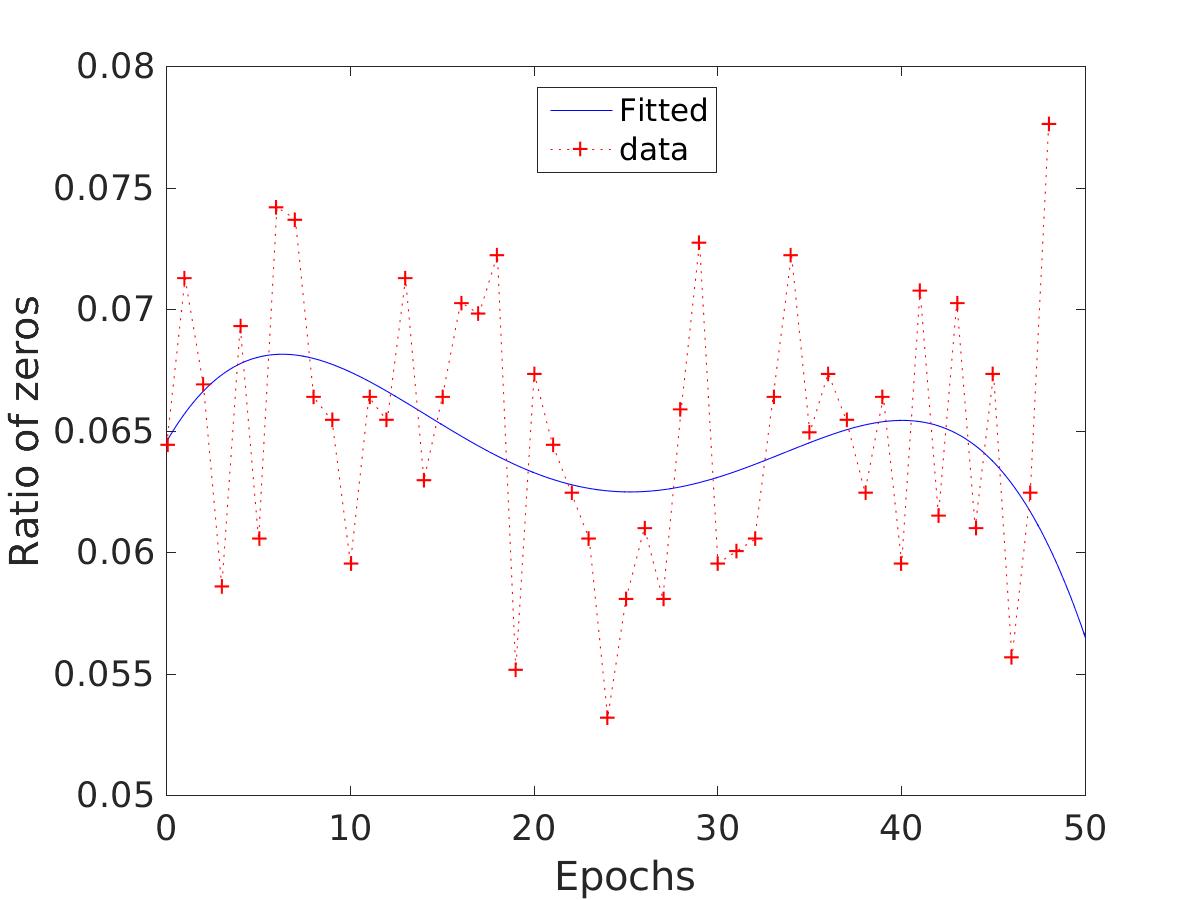} %{fig/H1-fit4-crop.pdf}
  \caption{Ratio of zero entries}
\label{fig:epoch-sparse}
\end{subfigure}%
\caption{Sensitivity and Optimization}
\end{figure*}

We analyze the optimization performance of SCoNet varying with training epochs. (One epoch means that the algorithm goes through the whole training dataset one time.) Results are shown on the Mobile dataset only due to space limit and the trend on the Amazon dataset is similar.

We firstly show the training loss and test performance. Figure~\ref{fig:epoch-loss} shows the training loss (averaged/normalized over all training examples) and NDCG test performance on the test set (HR and MRR have similar trends) varying with each optimization iteration. We can see that with more iterations, the training losses gradually decrease and the recommendation performance is improved accordingly. The most effective updates are occurred in the first 15 iterations, and performance gradually improves until 30 iterations. With more iterations, SCoNet is relatively stable.

Since the sparsity of transfer matrices $(\bm{H}^l)_1^L$ is crucial to select representations for transferring, we show the change of zero entries over training epochs. For clarity and due to space limit, we only show the results of the first transfer matrix $\mathbf{H}^1$ which connects the first and the second hidden layers. Figure~\ref{fig:epoch-sparse} shows the results where we use a 4-order polynomial to robustly fit the data. We can see that the matrix becomes sparser for the first 25 iterations, and the general trend is to sparsify. The average percent of zero entries in $\mathbf{H}^1$ is 6.5\%. For the second and third transfer matrices, the percentage becomes 6.0\% and 6.3\%, respectively. In summary, sparse transfer matrices are learned and they can adaptively select partial representations to transfer across domains. And it may be better to {\it transfer many instead of all} representations at hand.

For the training time, our models spend about 100 seconds per epoch using one Nvidia TITAN Xp GPU. As a reference, it is 70s for MLP and 90s for CSN, which indicates that the training cost of the proposed method is comparable to (non-)transfer deep baselines.

\section{Related Works}\label{related-work}

%Our work is mainly related to two research fields.

\noindent
{\bf Recommender systems } Recommender systems aim at learning user preferences on unknown items from their past history. Content-based recommendations are based on the matching between user profiles and item descriptions~\cite{CBF}. It is difficult to build the profile for each user when there is no/few content. Collaborative filtering (CF) alleviates this issue by predicting user preferences based on the user-item interaction behavior, agnostic to the content~\cite{IBCF}. Latent factor models learn feature vectors for users and items mainly based on matrix factorization (MF)~\cite{koren:2009} which has probabilistic interpretations~\cite{PMF,GeoMF}. MF is also flexible to integrate text~\cite{TBPR,kim2016convMF}, social relations~\cite{MR3,yang2017bridging}, and implicit feedback~\cite{koren:2009,MR3PP}. Factorization machines can mimic MF~\cite{rendle:2012}. Some hierarchical methods can reduce to factorize a specific matrix~\cite{HRM15}. Random walk and heterogeneous networks are adapted for recommendation~\cite{tempRW,shi2015semantic}. Neural networks are proposed to push the learning of feature vectors towards non-linear representations~\cite{dziugaite2015neural,DeepYoutube,he2017neural}. CF models, however, suffer from the data sparsity issue.

Cross-domain recommendation~\cite{cantador2015cross} is an effective technique to alleviate sparse issue. A class of methods are based on MF applied to each domain, including collective MF (CMF)~\cite{singh:2008} with its heterogeneous variants~\cite{pan2011transfer} and codebook transfer~\cite{li2009can,li2011cross}. Active learning~\cite{active-cross} can construct entity correspondence with limited budget. Heterogeneous cross-domain~\cite{heteroCross} and multiple source domains~\cite{lu2013selective} are also proposed to account for different cases of input. These are all shallow methods and have the difficulty in learning complex (highly nonlinear) user-item interaction relationship~\cite{DeepYoutube,he2017neural,RNNNetEase}. We follow this research thread by using deep networks to learn the nonlinear interaction function.

\noindent
{\bf Transfer and multitask learning } Transfer learning (TL) aims at improving the performance of the target domain by exploiting knowledge from source domains~\cite{pan2010survey}. The typical TL technique in neural networks is two-step: initialize a target network with transferred features from a pre-trained source network~\cite{mid_transfer,yosinski2014transferable}. Different from this approach, we transfer knowledge in a deep way such that two base networks benefit from each other during the learning procedure. Similar to TL, the multitask learning (MTL) is to leverage useful knowledge in multiple related tasks to help each other~\cite{MTL,zhang2017survey}. Multi-view learning~\cite{MVCross15} is closely related to MTL. The cross-stitch network (CSN)~\cite{misra2016cross} enables information sharing between two base networks. We generalize CSN by relaxing the underlying assumption, especially via the idea of selecting representations to transfer.

\section{Conclusions}\label{paper:conclusion}

We proposed a novel deep transfer learning for cross-domain recommendation. The sparse target user-item interaction matrix can be reconstructed with the knowledge guidance from the source domain, alleviating the data sparse issue. We demonstrated the necessity of adaptively selecting representations from the auxiliary domain to transfer. It may harm the performance by transferring all of them with equal importance. We found that naive deep transfer models may be inferior to the shallow/neural non-transfer methods in some cases. Our transfer model can reduce tens of thousands training examples by comparing with the non-transfer methods without performance degradation. This is useful when collecting data is difficult or costly. Experiments demonstrate the effectiveness of the proposed models on two large real-world datasets by comparing with shallow/deep, single/cross-domain methods. As a future work, we will integrate content information into the collaborative cross network for alleviating the cold-start problem.

%\newpage
%\section*{Acknowledgment}
{\bf Acknowledgment} We thank Cheetah Mobile for providing the Mobile dataset. The research has been supported by National Grant Fundamental Research (973 Program) of China under Project 2014CB340304, Hong Kong CERG projects 16211214/16209715/16244616, and NSFC 61673202. The work is also supported by HKPFS PF15-16701.

\bibliographystyle{ACM-Reference-Format}
\bibliography{bibliography}
	
\end{document}